\documentclass[11pt,a4paper]{article}
\usepackage[utf8]{inputenc}
\usepackage{fix-cm}
\usepackage{newunicodechar}
\newunicodechar{・}{\textperiodcentered}
\newunicodechar{－}{-}      
\newunicodechar{−}{-}     
\newunicodechar{‐}{-}     
\newunicodechar{–}{--}    
\newunicodechar{é}{\'{e}} 

\usepackage{graphicx}
\usepackage{subcaption}

\usepackage{amsmath}
\usepackage{mathrsfs}
\usepackage{xcolor}
\usepackage{soul}
\usepackage{siunitx}
\usepackage{framed} 
\usepackage{bm}
\usepackage{setspace} 
\usepackage{natbib} 
\usepackage{booktabs}
\usepackage{makecell}
\usepackage{tabularx}
\usepackage[left=3cm,right=3cm,top=3cm,bottom=3cm]{geometry}
\usepackage{empheq}
\usepackage{url}
\usepackage{authblk} 
\renewcommand{\textbf}[1]{#1}
\renewcommand{\bm}[1]{#1}

\newcommand{\keywords}[1]{\par\vspace{0.5em}\noindent\textit{Keywords:} #1}

\begin{document}

\title{Mathematical Modeling of Lesion Pattern Formation in Dendritic Keratitis
}

\author[1,2]{Mari Masunaga}
\author[3]{Reo Shimatani}
\author[4]{Kazumi Shinozaki}
\author[4,5]{Tomohiro Iida}
\author[2]{Yoshinao Oda}
\author[1]{Takashi Miura}

\affil[1]{Department of Anatomy and Cell Biology, Graduate School of Medical Sciences, Kyushu University, Fukuoka, Japan. Tel.: +81-92-6426048; Fax: +81-92-678910. Email: miura.takashi.869@m.kyushu-u.ac.jp}
\affil[2]{Department of Anatomic Pathology, Graduate School of Medical Sciences, Kyushu University, Fukuoka, Japan}
\affil[3]{Medical Institute of Bioregulation, Graduate School of Medical Sciences, Kyushu University, Fukuoka, Japan}
\affil[4]{Department of Ophthalmology, Tokyo Women's Medical University School of Medicine, Tokyo, Japan}
\affil[5]{Ageo Central General Hospital, Saitama, Japan}

\date{}

\maketitle

\begin{abstract}
Dendritic keratitis is a form of eye infection caused by herpes simplex virus (HSV). The virus spreads via direct cell-to-cell infection among corneal epithelial cells. This leads to the formation of dendritic lesions characterized by terminal bulbs at their tips. Under immunosuppression, the condition may progress to geographic keratitis, which is a map-shaped lesion with dendritic tails. The mechanism of this pattern formation remains to be elucidated. In this study, we propose a mathematical model to elucidate the mechanisms of lesion pattern formation in dendritic keratitis. Our model shows that increased production of infection-suppressive cytokines induces dendritic patterns with terminal bulbs, whereas reduced cytokine levels lead to geographic patterns. Furthermore, altering the spatial distribution of cytokine production can reproduce dendritic tails. By including external cytokine secretion, we could reproduce tapered lesions observed in non-HSV keratitis. By clarifying the mechanisms behind terminal bulb formation and reproducing atypical lesion morphologies, our findings enhance the understanding of herpetic keratitis and highlight the utility of mathematical modeling in ophthalmology.

\keywords{Dendritic keratitis \and Herpes simplex virus \and Mathematical modeling \and Branch pattern formation}
\end{abstract}

\newpage
\doublespacing

\section{Introduction}
\label{sec:Introduction}
\subsection{Biological Background}
\subsubsection{Herpes Keratitis}
\label{subsubsec:Herpes_Keratitis}
Several ocular diseases are characterized by corneal epithelial defects with a branching pattern reminiscent of tree branches. Among these, the most representative is dendritic keratitis, an infectious epithelial keratitis caused by herpes simplex virus (HSV) replication in the corneal epithelium, the most superficial layer of the cornea \cite{MarkJMannis2021}.
The seroprevalence of HSV worldwide is approximately 90\% \cite{MarkJMannis2021}. Dendritic keratitis can result from direct viral entry into the ocular surface via droplet transmission \cite{Kaye2006}. However, in most cases, individuals acquire HSV infection during childhood, and the virus remains latent until reactivation occurs due to immunosuppression or other triggering factors, leading to the development of dendritic keratitis \cite{MarkJMannis2021}.

\subsubsection{Keratitis with branching lesions}
\begin{figure}[htbp]
    \centering
        \centering
        \includegraphics[width=\linewidth]{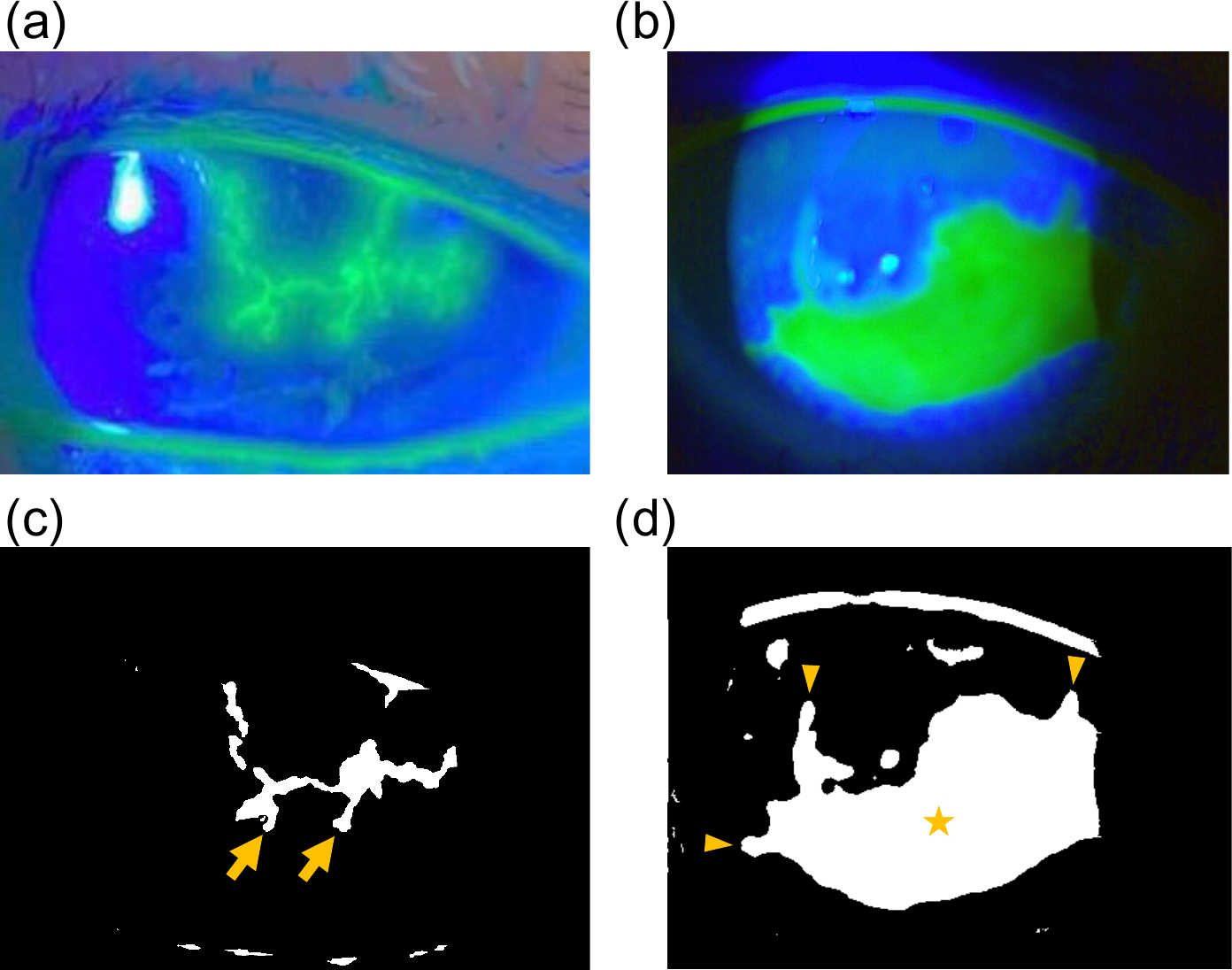}
    \hfill
    \caption{\textbf{Slit-lamp photographs of dendritic keratitis (a) and geographic keratitis (b), and their corresponding binary images (c, d).
    Panels (a) and (b) were taken under cobalt-blue illumination after fluorescein staining.
    Fluorescein dye stains corneal regions where cell-to-cell junctions are disrupted, and visualize diseased epithelial defects \cite{Srinivas2023}.
    As a result, the ulcerated epithelial areas appear green, while the surrounding intact corneal surface appears blue. Panels (c) and (d) show binary images generated from (a) and (b), respectively, using the Color Threshold function in ImageJ to enhance lesion visibility.
    (a)(c) The lesion exhibits a characteristic branching, linear ulcerative pattern with terminal bulbs (arrows) and swollen epithelial borders. 
    (b)(d) The lesion exhibits a broad amoeboid ulcerative pattern (stars). At the lesion periphery, dendritic tails (arrowheads) are observed.}} 
    \label{fig:fig_keratitis_samples}
\end{figure}

\paragraph{Dendritic Keratitis}
The features of dendritic keratitis include a branching, linear ulcerative lesion with terminal bulbs and swollen epithelial borders (Fig. \ref{fig:fig_keratitis_samples}a\textbf{, \ref{fig:fig_keratitis_samples}c, Supplementary Figure S1}) that contain live virus \cite{MarkJMannis2021,Azher2017}.
In immunocompromised patients or those receiving corticosteroid therapy, the lesions may progress to geographic keratitis \cite{Prakash2015, Wilhelmus2015}, sometimes with dendritic tails \cite{Azher2017} (Fig. \ref{fig:fig_keratitis_samples}b\textbf{, \ref{fig:fig_keratitis_samples}d}).

The morphogenesis of dendritic keratitis is thought to be closely related to viral infection and the host immune response 

\textbf{\cite{Thakkar2017}}. The virus spreads through a cell-to-cell infection mechanism, where viral particles are directly transmitted between adjacent corneal epithelial cells \textbf{\cite{Kaye2006}}. 
\textbf{Consistent with this, dendritic spread occurred in organotypically cultured corneas but not in isolated corneal cell cultures \cite{Thakkar2017}, suggesting that cell-to-cell transmission is essential for the formation of dendritic lesions.} 

Infected cells undergo nuclear expansion due to viral replication, forming lesions within 12–24 hours \cite{Kaye2006}. Since the cornea is an avascular tissue, early viral clearance primarily depends on the innate immune response \cite{Rowe2013}. Additionally, in vitro or ex vivo studies have demonstrated that infected corneal epithelial cells secrete antiviral cytokines such as interferon-alpha (IFN$\alpha$) and interferon-beta (IFN$\beta$), which play a role in suppressing infection \cite{Royer2016, Thakkar2017}.

\textbf{Though the dendritic pattern of dendritic keratitis was once thought to correspond to the network of intraepithelial corneal nerves, subsequent studies using a rabbit in vivo model \cite{Baum1970} and a human ex vivo corneal model \cite{Courrier2020} have demonstrated that the morphology of dendritic keratitis is independent of corneal nerve distribution.}

\begin{figure}[htbp]
  \centering
  \includegraphics[width=0.45\linewidth]{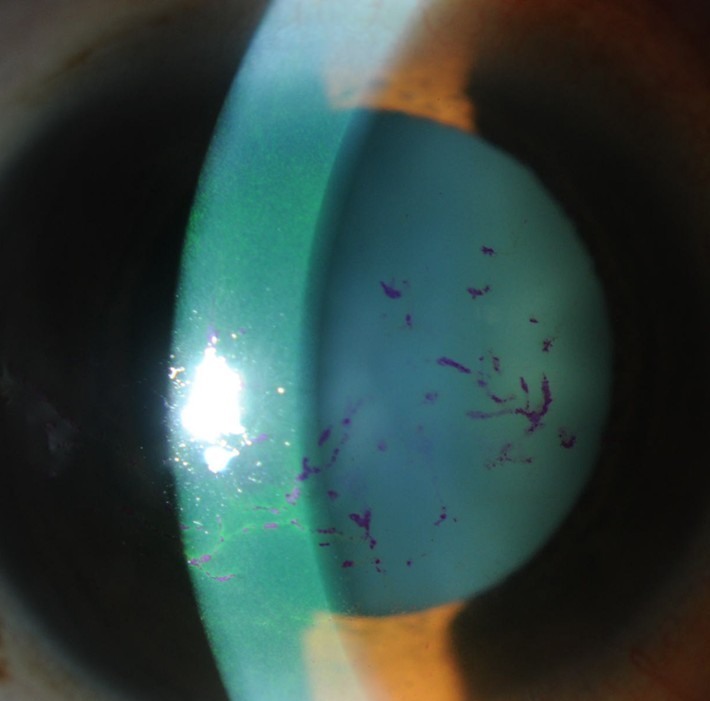}
  \caption{Slit-lamp photographs of pseudodendritic keratitis of VZV infection. 
  Pseudodendritic keratitis lacks terminal bulbs, and tends to form thinner lesions.
  }
  \label{fig:fig_VZV}
\end{figure}

\paragraph{Pseudodendritic Keratitis}
Keratitis caused by non-HSV pathogens for example, varicella-zoster virus (VZV) or Acanthamoeba can also exhibit branching lesions.
Such keratitides not caused by HSV are referred to as pseudodendritic keratitis \cite{Jain2006}.
Compared to dendritic keratitis, pseudodendritic keratitis lacks terminal bulbs, and tends to form thinner lesions \cite{Suzuki2023} \textbf{(Fig.~\ref{fig:fig_VZV})}.
Pseudodendritic keratitis caused by VZV, in particular, is accompanied by conjunctivitis
\cite{Suzuki2023}.

\paragraph{Differential Diagnosis and Treatment}
The diagnosis of dendritic keratitis is primarily based on clinical findings, including patient history and morphological assessment of the lesion \cite{Hoarau2023}. 
Since pseudodendritic keratitis resembles dendritic keratitis, careful diagnosis is essential.
Misdiagnosis can occur in both directions, dendritic keratitis may be mistaken for pseudodendritic keratitis \cite{Singh2018}, 
and conversely, pseudodendritic keratitis may be misdiagnosed as dendritic keratitis \cite{Lee2020}. 

A study on clinically diagnosed herpes keratitis cases revealed that 5\% of these lesions were actually caused by adenovirus, while 2.7\% were due to enterovirus \cite{Marangon2007}, highlighting the diagnostic challenges associated with this condition.
Dendritic keratitis may resolve spontaneously, and antiviral treatment is generally effective. However, delays in the diagnosis and appropriate treatment of HSV keratitis can lead to severe complications including perforation, sometimes resulting in the need for corneal transplantation \cite{Zhou2024}. 
Conversely, failure to recognize non-HSV etiologies, such as Acanthamoeba keratitis, can result in severe visual impairment, including blindness, if appropriate treatment is not initiated promptly \cite{Lorenzo-Morales2015}. 
Therefore, accurate diagnosis of dendritic keratitis is crucial. A comprehensive understanding of the morphogenesis of terminal bulbs, a key feature of dendritic keratitis, is essential for precise diagnosis.

\textbf{
However, the precise mechanisms underlying the morphogenesis of dendritic keratitis remain unclear.}
\textbf{This knowledge gap is even more pronounced for VZV keratitis, for which both lesion morphogenesis 
and fundamental aspects of its pathophysiology remain poorly characterized, largely due to the virus’s 
strict human specificity and the lack of good experimental models\cite{Kennedy2021}.}

\subsection{Mathematical Models of branch pattern formation}

\subsubsection{Models for Dendritic Patterns}
Several mathematical models have been proposed to reproduce the expansion of dendritic patterns, including the Diffusion-Limited Aggregation (DLA) model \cite{Witten1983} and the L-system \cite{Lindenmayer1968}. 
The DLA model is a probabilistic model in which particles undergo random walks and adhere to an existing structure upon contact, resulting in the formation of irregular fractal patterns. This model has been applied in biological contexts, such as the morphology of bacterial colonies \cite{Matsushita1990} and coral growth \cite{Kaandorp2013}.  
The L-system (Lindenmayer system) is a mathematical framework based on recursively applied production rules, generating hierarchical and self-similar patterns \cite{Lindenmayer1968}. It has been widely used to model branching structures in plants \cite{Prusinkiewicz1986, DeJong2022}.  

\subsubsection{Models for Geographic Patterns}
The Eden model \cite{Eden1961} is known for reproducing growth of geographic patterns. It is a stochastic growth model in which structures expand probabilistically, leading to irregular morphologies that follow scaling laws. Initially proposed as a model for proliferating cell populations, the Eden model has been applied in studies on bacterial colony formation \cite{Kaczmarczyk2025}.  

\subsubsection{Models Capable of Reproducing Both Dendritic and Geographic Patterns}
Some models can represent both dendritic and geographic patterns, such as the Mimura model \cite{Mimura2000} and the phase-field model \cite{Kobayashi1993}. The Mimura model is a reaction-diffusion system that describes bacterial growth driven by nutrient consumption, capable of generating a variety of patterns, including dendritic, Eden-type, circular, and concentric ring structures.  

The phase-field model was originally proposed to describe the movement of interfacial boundaries in crystal growth and represents this process using reaction-diffusion equations \cite{Kobayashi1993}. It can capture both uniform growth and interfacial instabilities. While this model has been widely utilized in materials science \cite{Tourret2022}, it has also been increasingly applied to biological pattern formation in recent years \cite{Nonomura2012, Miura2009}. These models typically produce smooth and regular patterns due to the influence of diffusion terms.

\subsubsection{Interfacial Instabilities}
In many models, branching structures develop through a positive feedback mechanism, where protruding regions grow faster. For example, in bacterial colonies, the leading edges of the protrusions are exposed to higher nutrient concentrations, leading to accelerated growth \cite{Miura2008}.  

\subsubsection{Application to dendritic Keratitis}
Although these models effectively describe various branching and expansion processes, they have not been applied to the morphogenesis of dendritic lesions of dendritic keratitis.

\subsection{Summary}
In this study, we aimed to clarify the mechanism underlying the formation of dendritic keratitis, using mathematical models. To implement the cell-to-cell infection process, we developed a probabilistic mathematical model in which uninfected cells adjacent to infected cells stochastically undergo transition into an infected state. Additionally, cytokines with infection-suppressive effects were assumed to be produced by infected cells or neighboring uninfected cells, reducing the probability of infection.

The validity of the constructed mathematical model was evaluated through numerical simulations and image analysis, comparing the results with clinical images. Furthermore, dendritic keratitis without terminal bulbs were successfully reproduced by incorporating the effects of externally derived cytokines rather than direct infection of cells. These findings indicate that mathematical modeling is a valuable tool for understanding the morphogenesis of dendritic keratitis.

\section{Models}
\label{Models}
In this study, we developed mathematical models to describe the infection dynamics of cells, that are expressed as discrete square lattices. 
To emulate cell-to-cell infection, our models assume that uninfected cells adjacent to infected cells can probabilistically become infected.
The spread of infection is negatively correlated with the concentration of cytokines, which act as infection inhibitors.
The secretion, degradation, and diffusion of cytokines are described using a reaction-diffusion equation.

\subsection{Model for the Formation of Dendritic Keratitis}
\label{subsec:Model_dendritic_keratitis}
As confirmed in in vitro and ex vivo studies using infected cells \cite{Thakkar2017, Royer2016}, our models assume that infection-suppressing cytokines are produced by infected cells.
Additionally, since it is possible that cytokines are also produced by uninfected cells in close proximity to infected cells like cancer-associated fibroblasts (CAFs) \cite{Sahai2020}, we developed models that consider cytokine secretion from uninfected cells as well.

\subsubsection{Cell Infection Dynamics}
\label{subsubsec:Cell_Infection_Dynamics}
We defined $u_{i,j}(t)$ as the \textbf{dimensionless}state variable representing whether a cell at position $(i,j)$ at time $t$ is infected ($u_{i,j} = 1$) or uninfected ($u_{i,j} = 0$) (Fig. \ref{fig:fig_variables}a). The update rule for $u_{i,j}(t)$ is given by:
\begin{equation}
\bm{
    u_{i,j}(t+dt)
    = u_{i,j}(t)
    + B_{i,j}(t) 
    H\!\bigl(
        \,\eta_{i,j}(t) <
        dt ( \alpha- \beta\,v_{i,j}(t) )
      \bigr),
}
\label{eq:u}
\end{equation}

\textbf{where} $B_{i,j}(t)$ is a \textbf{dimensionless}indicator variable that determines whether the cell at $(i,j)$ is  \textbf{uninfected and adjacent to at least one infected cell}:
\begin{equation}
\bm{
B_{i,j}(t)=\bigl(1 - u_{i,j}(t)\bigr)
  \min\!\Bigl(
    1,\;
    u_{i-1,j}(t)
  + u_{i+1,j}(t)
  + u_{i,j-1}(t)
  + u_{i,j+1}(t)
  \Bigr).
}
\label{eq:B}
\end{equation}

\textbf{
\bm{$H(x)$} denotes the Heaviside step function, defined as 
\bm{$H(x) = 1$} if \bm{$x > 0$} and \bm{$H(x) = 0$} otherwise.
}

\textbf{
The dimensionless parameter} \bm{$\eta_{i,j}(t)$} \textbf{is an independently sampled random number
drawn from a uniform distribution on \bm{$(0,1)$} at each lattice site \bm{$(i,j)$} and each time
step \bm{$dt$}. Infection is assumed to occur during a time interval \bm{$dt$} when
\bm{$\eta_{i,j}(t)$} is smaller than \bm{$dt(\alpha-\beta v_{i,j}(t))$},
thereby implementing a stochastic infection process.}
\textbf{The parameter $\alpha$ represents a cytokine-independent net infection drive per unit time, incorporating intrinsic viral infectivity (e.g., baseline viral entry efficiency) together with static protective mechanisms, such as physical barriers (e.g., extracellular matrix or tight junctions).}
 The parameter $\beta$ is the infection inhibition coefficient influenced by cytokines, corresponding to cellular sensitivity to cytokine signaling.

\subsubsection{Cytokine Dynamics}
\label{subsubsec:Cytokine_Dynamics}
We defined $v_{i,j}(t)$ as the cytokine concentration at position $(i,j)$ at time $t$. \textbf{The unit of 
$v_{i,j}(t)$ is an arbitrary reference concentration denoted by [C$_0$]}. The evolution of cytokine concentration follows the equation:
\begin{equation}
    v_{i,j}(t+\bm{dt}) = 
    \bm{
    v_{i,j}(t)
    +
    dt}
    \Big(
    \gamma S_{i,j}(t) - \delta v_{i,j}(t) +
    D_v \Delta v_{i,j}
    \bm{
    (t+dt)
    }
    \Big).
\label{eq:v}
\end{equation}

\begin{figure}[htbp]
  \centering
  \includegraphics[width=0.95\linewidth]{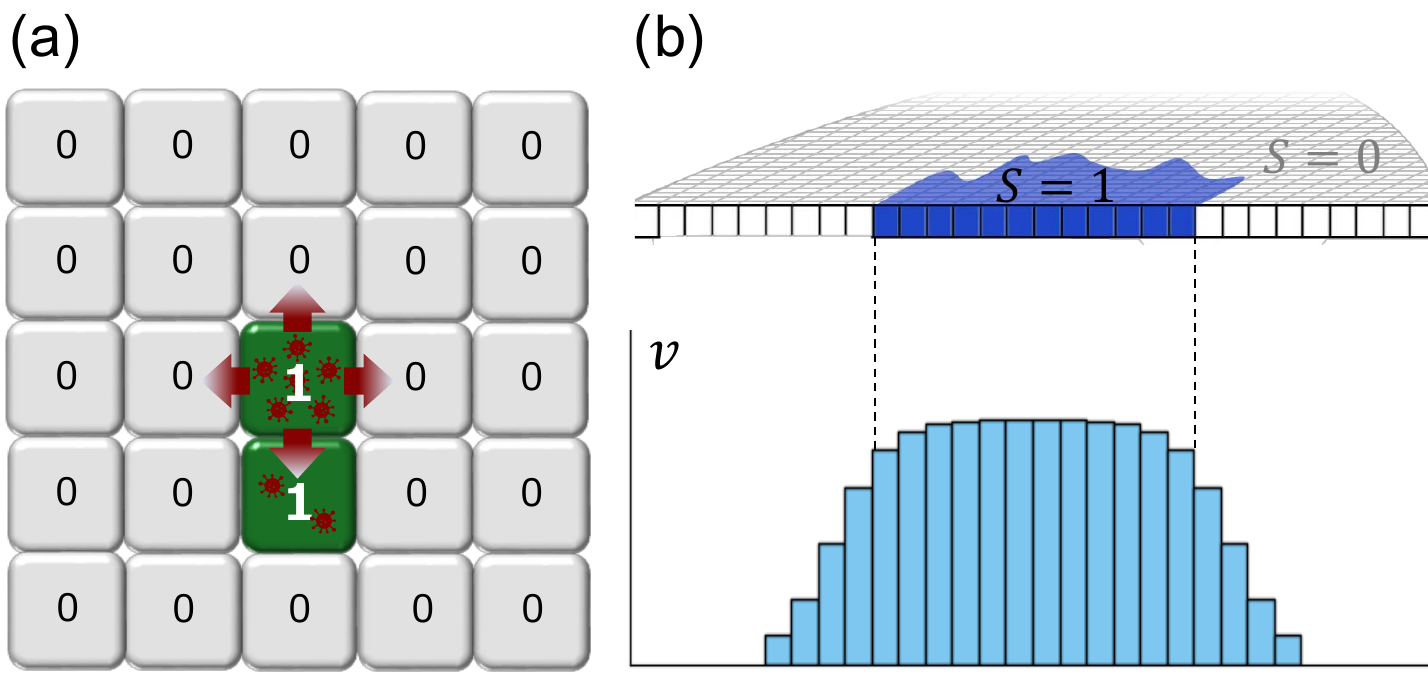}
  \caption{Conceptual diagrams of variable $u$,  $S$ and $v$. (a) Conceptual diagram of variable $u$.
        We defined $u=1$ for infected cells and $u=0$ for uninfected cells.
        Uninfected cells ($u=0$) adjacent to infected cells ($u=1$) can probabilistically become infected cells ($u=1$).
        (b) Conceptual diagram of variable $S$ and $v$.
    Cytokines are produced from the cytokine-producing area where $S=1$.
    The cytokine concentration $v$ is high within the region ($S=1$), and spreads to the surrounding area through diffusion.    
  }
  \label{fig:fig_variables}
\end{figure}

\textbf{
Here, the diffusion term is evaluated at $\bm{t+dt}$ because the diffusion process was computed using an implicit scheme.
}
The parameter $\gamma$ is the cytokine secretion rate, representing the level of cytokine production.
The parameter $\delta$ is the cytokine degradation rate, corresponding to the rate at which cytokines are cleared from the local environment.
The parameter $D_v$ represents the diffusion coefficient of cytokines, reflecting the ability of cytokines to spread spatially through the tissue. 
The variable $S_{i,j}(t)$ is a binary indicator that represents whether a cell at $(i,j)$ is secreting cytokines (Fig. \ref{fig:fig_variables}b).

The diffusion term $\Delta v_{i,j}(t)$ is given by:
\begin{equation}
    \Delta v_{i,j}(t) = v_{i-1,j}(t) + v_{i+1,j}(t) + v_{i,j-1}(t) + v_{i,j+1}(t) - 4 v_{i,j}(t).
\label{eq:lap}
\end{equation}
\textbf{The diffusion was discretized using a standard four-neighbor finite-difference scheme in two dimensions. Diagonal interactions were not included for simplicity.}This term represents the local spatial dispersal of cytokines. 

\begin{figure}[htbp]
  \centering
  \includegraphics[width=0.8\linewidth]{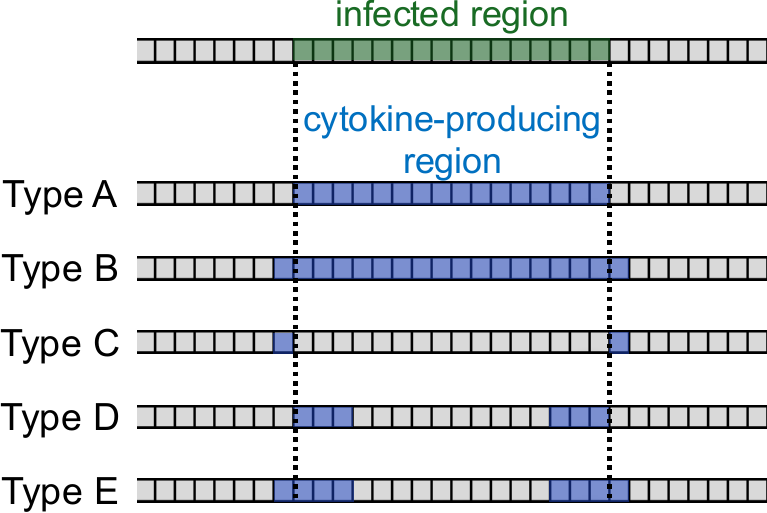}
  \caption{
    Conceptual diagram of the infected region (green) and the cytokine-producing region (blue).
    The top row represents the infected region (green), while Types A to E show the cytokine-producing regions (blue) in each case.
    The dashed lines indicate the boundaries of the infected region.
    The five cases are as follows:
    (Type A) All infected cells produce cytokines.
    (Type B) All infected cells and their adjacent uninfected cells produce cytokines.
    (Type C) Only adjacent uninfected cells produce cytokines.
    (Type D) Only infected cells located within $n$ layers of the lesion boundary produce cytokines.
    (Type E) Both infected cells located within $n$ layers of the lesion boundary and their adjacent uninfected cells produce cytokines.
  }
  \label{fig:fig_S}
\end{figure}

We considered the following five cases for defining the cytokine-producing regions, represented by the \textbf{dimensionless}variable $S_{i,j}(t)$. 
These five cases reflect biologically plausible scenarios of cytokine production.
The five cases are as follows (Fig. \ref{fig:fig_S}):

(Type A) all infected cells produce cytokines;
\begin{equation}
S_{i,j}(t) = u_{i,j}(t).
\label{eq:S_A}
\end{equation}

(Type B) All infected cells and their adjacent uninfected cells produce cytokines;
\begin{equation}
\bm{S_{i,j}(t) = u_{i,j}(t) + B_{i,j}(t).}
\label{eq:S_B}
\end{equation}

(Type C) Only adjacent uninfected cells produce cytokines;
\begin{equation}
\bm{
S_{i,j}(t) = B_{i,j}(t).
}
\label{eq:S_C}
\end{equation}

(Type D) Only infected cells located within $n$ layers of the lesion boundary produce cytokines;

\begin{equation}
\bm{
S_{i,j}(t) = B^{(n)}_{i,j}(t),
}
\label{eq:S_D}
\end{equation}
\textbf{where}
\begin{equation}
\bm{
B^{(n)}_{i,j}(t) = 
u_{i,j}(t)
\min\!\left(1,\; \sum_{1 \le |k|+|l| \le n} (1 - u_{i+k,j+l}(t)) \right).
}
\label{eq:S_D_Bn}
\end{equation}

(Type E) Both infected cells located within $n$ layers of the lesion boundary and their adjacent uninfected cells produce cytokines;
\begin{equation}
\bm{
S_{i,j}(t) =
B_{i,j}(t)+
B^{(n)}_{i,j}(t).
}
\label{eq:S_E}
\end{equation}

\subsection{Modified Model for Formation of Pseudodendritic Keratitis}
\label{subsec:modified_model}
As stated in the Introduction, dendritic keratitis typically presents with terminal bulbs at the distal ends of the lesions, whereas pseudodendritic keratitis lacks terminal bulbs and tends to form thinner lesions.
As also mentioned in the Introduction, pseudodendritic keratitis by VZV is accompanied by conjunctivitis.
This pronounced clinical observation in non-HSV infections suggests the host immune response may act more strongly in VZV infections than in HSV infections, thereby restricting the spread of infection more effectively.

To gain a better understanding of the mechanism underlying the morphological differences between dendritic keratitis and pseudodendritic keratitis, we modified the above-mentioned model to incorporate the effects of strong host immune response.

\begin{figure}[htbp]
  \centering
  \includegraphics[width=0.8\linewidth]{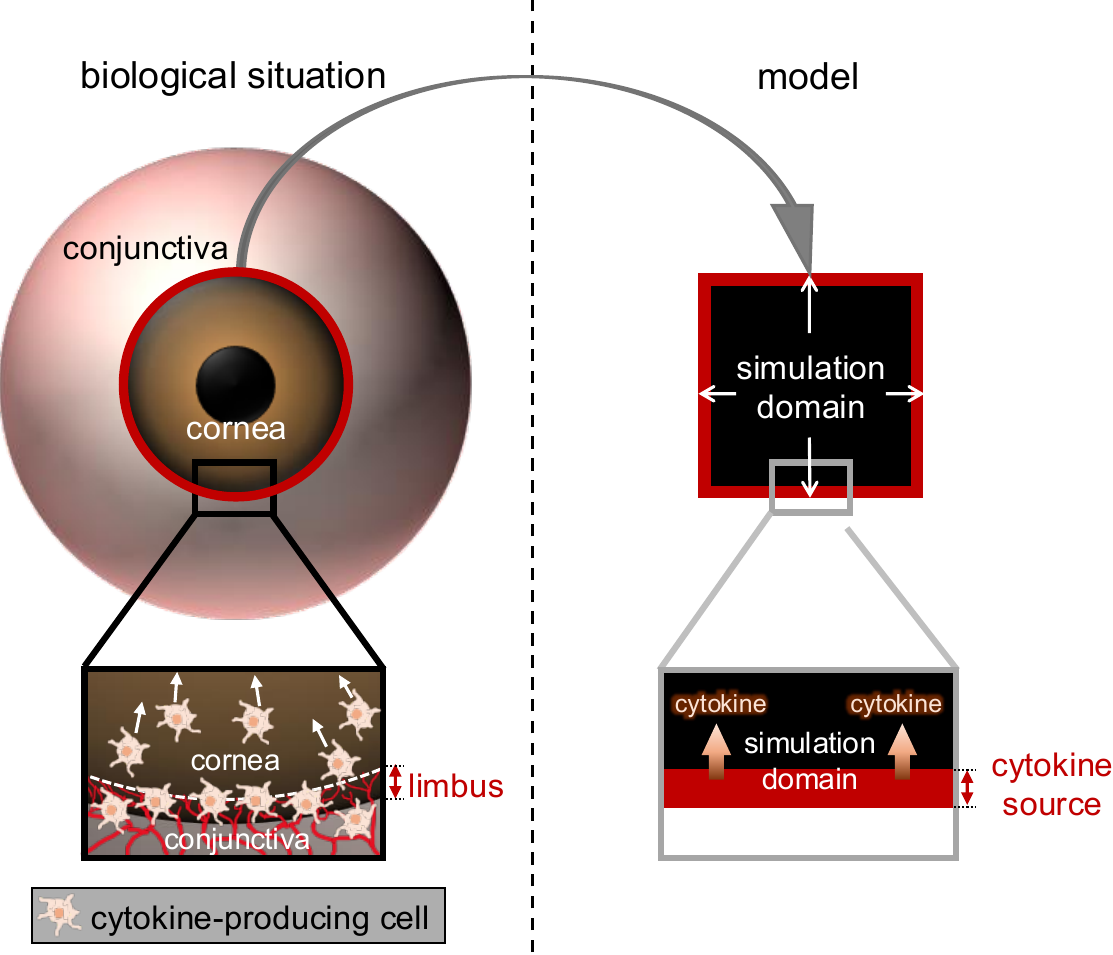}
  \caption{Conceptual diagram of simulation domain under inflammatory conditions.
  \textbf{During inflammation, cytokine-producing cells (e.g., dendritic cells) migrate from the limbus into the cornea and locally release cytokines. In the numerical model, this process is approximated by placing cytokine-producing source regions along the boundary of the simulation domain.}
  }
  \label{fig:fig_VZV_model}
\end{figure}

Although the cornea is normally avascular, its peripheral margin, known as the limbus, contains a rich vascular network 
\cite{Schlotzer-Schrehardt2005, Li2007}.
Immune cells and inflammatory substances can enter the cornea from the limbal blood vessels during inflammation \cite{Mousa2021}. 
\textbf{
Indeed, experimental studies in mouse corneas have shown that dendritic cells are sparse in the central cornea under non-inflamed conditions, but their density increases in the central region when inflammation is induced \cite{Hamrah2003}. An increase in dendritic cell density during inflammation is also known to correlate with elevated cytokine levels \cite{Yamaguchi2016}. These findings collectively suggest that cytokine-producing immune cells located at the limbus can be recruited toward the lesion site in the central cornea during inflammation. 
Therefore, in this modified model, we approximate this recruitment process by allowing infection-suppressive cytokines to originate not only from infected cells but also from source regions placed along the limbal boundary, from which they diffuse toward the central cornea} (Fig.\ref{fig:fig_VZV_model}).

\textbf{As described in Section~\ref{subsec:method_numerical_simulation},
the simulation was conducted on a grid of size $1000 \times 1000$, indexed from $(0,0)$ to $(999,999)$.
All grid points were included in the computational domain.
The outermost layer of cells, defined by $i=0$, $i=999$, $j=0$, or $j=999$, was designated as the cytokine source region and released cytokines at a constant rate $C$.}
A new term \bm{$w([C_0])$}representing cytokine supply from \textbf{limbus} was incorporated into the update equation for cytokine concentration $v$.
\textbf{Furthermore, because this process represents the movement of immune cells from the limbus toward the lesion site, we modeled it using an external cytokine component with a relatively large effective diffusion coefficient $D_w$, allowing for an effectively long diffusion length. The effective diffusion length is given by $\lambda = \sqrt{\frac{D_w}{\delta}}$, and using the parameter values specified in the Material and methods section, it is approximately 10 mm. This long diffusion length is consistent with biological observations that central corneal lesions increase dendritic cell density even at substantial distances from the limbus \cite{Hamrah2003}.}

\begin{align}
v_{i,j}(t+\bm{dt}) &= 
\bm{v_{i,j}(t)
+
dt}
\Big(
\gamma S_{i,j}(t) - \delta v_{i,j}(t)  
+ D_v \Delta v_{i,j}(t+\bm{dt})
\Big)
 + w_{i,j}(t+\bm{dt})
, \\
w_{i,j}(t+\bm{dt})
&= 
\begin{cases}
\bm{
w_{i,j}(t)
+
dt}
\Big(
C + D_w \Delta w_{i,j}(t+\bm{dt})
\Big)
& \text{if } (i = 0 \text{ or } -1) \text{ or } (j = 0 \text{ or } -1), \\
\bm{w_{i,j}(t)
+
dt}
\Big(
D_w \Delta w_{i,j}(t+\bm{dt})\Big)
& \text{otherwise}.
\end{cases}
\label{eq:VZV}
\end{align}

\section{Materials and methods}
\subsection{Numerical simulation}
\label{subsec:method_numerical_simulation}
We performed numerical simulations using the finite difference method with periodic boundary conditions.

The domain size was set to $12 \times 12$ \si{mm}, based on the fact that the human cornea typically has a diameter of approximately $11$–$12$ mm \cite{Rufer2005}.
The cell size was set to $\Delta x = 0.012$ \si{mm}, referring to previous confocal microscopy observations of corneal epithelial cells \cite{Prakasam2013}.

\textbf{
Accordingly, the computational lattice consisted of $1000 \times 1000$ grid size.
}

The time step was set to $\bm{dt = 1.0}$. 

Based on visual similarity between experimental data from an ex vivo porcine cornea model \cite{Thakkar2017} and our simulation results,  
we estimated that 400 simulation steps (with a time step size of $ \Delta t = \bm{1.0} $) approximately correspond to 96 hours in real time.  
Accordingly, one simulation time step size nearly equals 0.5 hours.  

The parameter $\gamma$ was varied from $0.5$ to $10.0$ in increments of $0.5$.  
Other parameters were fixed as $\alpha = 0.5$, $\beta = 1.0$, $\delta = 0.5$, and $D_v = 0.05$.

In a modified model for the formation of pseudodendritic keratitis, the parameter $C$ was varied from \bm{$0.01$} to \bm{$0.10$} in increments of \bm{$0.01$}.
The parameter $D_w$ was set to 50.
\textbf{A summary of parameters used in this model is provided in Table~\ref{tab:simulation_parameters}.}

We initially explored a broader range of parameter values for all parameters, as shown in Fig. \ref{fig:other_parameters}, and subsequently narrowed it to include morphologies similar to those observed in clinical cases.

As the initial condition, only the single central cell of the domain was set to the infected state ($u = 1$), while all other cells were set to the uninfected state ($u = 0$).  
The initial cytokine concentration $v$ was set uniformly to zero over the whole domain.

\begin{table}[htbp]
\centering
\caption{List of variables and parameters used in the simulation}
\begin{tabularx}{\textwidth}{
    >{\hsize=0.3\hsize}X
    >{\hsize=0.35\hsize}X
    >{\hsize=0.3\hsize\footnotesize}X
    >{\hsize=1.5\hsize}X
}
\toprule
\textbf{Parameter} & \textbf{Value} & \textbf{Unit} & \textbf{Biological interpretation} \\
\midrule
\mbox{Domain size} & $12 \times 12$ & \textbf{mm} & Size of the human cornea~\cite{Rufer2005} \\

$\Delta x$ & $0.012$ & \textbf{mm} & Size of a corneal epithelial cell~\cite{Prakasam2013} \\

$dt$ & $\bm{1.0}$ & \textbf{h} & \textbf{$1$ simulation step} $\approx 0.5$ hours; estimated from visual similarity to ex vivo lesion data~\cite{Thakkar2017} \\

$\alpha$ & 0.5 & \textbf{1/h} & \textbf{Net infection drive coefficient (cytokine-independent; incorporating intrinsic infectivity, e.g., viral entry efficiency, and static protection, e.g., physical barriers)} \\

$\beta$ & 1.0 & $\bm{1/(\mathrm{h}\,[C_0])}$ & Infection inhibition coefficient (cytokine-dependent; e.g., cellular sensitivity to cytokines) \\

$\gamma$ & $0.5 \sim 10.0$ & \mbox{\textbf{[C$_0$]/h}} & Cytokine secretion rate \\

$\delta$ & 0.5 & \textbf{1/h} & Cytokine degradation rate \\

$D_v$ & 0.05 & \mbox{\textbf{mm$^{2}$/h}} & Diffusion coefficient of cytokines \\

$n$ & $1 \sim 10$ & \textbf{--} & The number of layers producing cytokine \\

$C$ & \mbox{$0.01 \sim 0.10$} & \mbox{\textbf{[C$_0$]/h}} & Secretion rate of cytokines at the limbus (modified model only) \\

$D_w$ & 50 & \mbox{\textbf{mm$^{2}$/h}} & \textbf{Effective diffusion coefficient representing immune-cell–derived cytokines} (modified model only) \\
\bottomrule
\end{tabularx}
\footnotesize{\vspace{2mm}
[C$_0$] denotes an arbitrary reference concentration used to nondimensionalize cytokine levels in the model.
}
\label{tab:simulation_parameters}\label{R1-4-table}
\end{table}

To ensure numerical stability, the cytokine concentration $v$ was monitored at each time step to verify that it remained non-negative ($v \geq 1 \times 10^{-10}$); rounding errors were considered negligible.

The maximum number of time steps was set to 10,000. 
However, the simulation was terminated earlier if the lesion expanded sufficiently in the model for the formation of dendritic keratitis.  
Specifically, the computation was stopped when infected cells ($u = 1$) were detected within 40\% of any boundary of the simulation domain, either vertically or horizontally,  
which corresponds to the lesion occupying approximately 20\% of the domain size.
On the other hand, in the case of the modified model for pseudodendritic keratitis, the computation was stopped when infected cells ($u = 1$) were detected within 10\% of any boundary of the simulation domain, either vertically or horizontally,  
which corresponds to the lesion occupying approximately 80\% of the domain size.

Simulations were implemented in Python 3.9.18, and numerical computations were carried out using the NumPy library.

\subsection{Visualization}
\label{subsec:visualization}
Visualization of the simulation results was performed using the \texttt{matplotlib.pyplot} library in Python.  
We generated binary images to represent the variable $u$, with uninfected cells ($u = 0$) shown in black and infected cells ($u = 1$) in white.  
When directly visualized as a grid, each cell corresponds to approximately 0.012 \si{mm} in actual size.
However, as noted in the Introduction, infected cells tend to expand due to nuclear expansion \cite{Kaye2006}, based on observation of images from previous studies in which HSV-1 was infected with human corneas ex vivo, infected cells appeared to be approximately 2–3 times larger than uninfected cells \cite{Courrier2020}.
Therefore, in this visualization, infected cells were plotted as circles using \texttt{plt.scatter}, with a diameter approximately 2.5 times larger than the grid size.
\textbf{Importantly, this enlargement was used exclusively for visualization and does not affect the infection dynamics.
In the numerical simulation, each grid point represents a single epithelial cell of fixed size, and all infection and spread rules (neighbor interactions, diffusion, and stochastic infection events) are computed on this fixed grid.
Because cellular swelling occurs only after infection as a degenerative change, it is not included in the mechanistic model and does not influence the formation of lesion morphology.}
To reduce minor irregularities and noise along the boundaries of the output images, smoothing was performed using the alpha shape algorithm with a parameter of $\alpha = 0.3$ \cite{Edelsbrunner1983}. 
This procedure was implemented using the \texttt{shapely} library and \texttt{scipy.spatial.Delaunay} \cite{Edelsbrunner1983}. 
After applying the alpha shape algorithm, any enclosed regions remaining inside the lesion were filled with white to complete the segmentation.

\subsection{Skeletonization and Graph-Based Structure Extraction}
Skeletonization was performed on the binary images obtained from the visualized numerical simulation results ((Section~\ref{subsec:visualization}, where infected regions are white and uninfected regions are black).
Skeletonization was applied using the \texttt{skimage.morphology.skeletonize} function.
On the skeleton, the point closest to the center of the image (i.e., the seed) was designated as the central node.
Starting from this central node, a graph structure was constructed by tracing adjacent pixels along the skeleton.
A depth-first search (DFS) algorithm was used to construct the graph structure.
From the constructed graph structure, each branch's path and its corresponding endpoint were identified.
Since the numerical simulations were initiated with a seed at the image center, only relatively long branches with endpoints at a certain Euclidean distance (at least 4\% of the image width) from the central node were retained for analysis.
Furthermore, for pairs of branches with paths overlapping by 50 pixels or more from the central node, the shorter one was excluded from the analysis.
To prevent the analysis results from being overly dependent on specific images, up to three longest branches were selected from each image. 

For each branch of each image, the distance map values (the shortest Euclidean distance from each skeleton pixel to the lesion boundary) were obtained along the skeleton from the central node to the endpoint using the \texttt{distance\_transform\_edt} function.
Since the terminal ends of the patterns are often close to the image boundary, the distance map values at these regions tend to be low. Therefore, if the distance map value at the terminal part of a branch was less than 3, the low-value tail was iteratively trimmed until a value of 3 or higher was reached. The remaining sequences of distance map values were used for further analysis. \textbf{The workflow used for branch selection and the number of branches retained at each filtering step are summarized in Supplementary Figure~S4.}

\subsection{Statistical Analysis of Terminal Bulb}

Even in the modified model for pseudodendritic keratitis model (Section~\ref{subsec:modified_model}), the influence of cytokines from the boundary of the computational domain (corresponding to the limbus) is considered minimal at the early stage of lesion expansion. 

Therefore, in both the models for dendritic keratitis (Section~\ref{subsec:Model_dendritic_keratitis}) and pseudodendritic keratitis (Section~\ref{subsec:modified_model}), the analysis was restricted to dendritic patterns whose central nodes had similar distance map values.

Specifically, in parameter settings that produced morphologies resembling those observed in clinical cases of pseudodendritic keratitis, the distance map values at the central nodes were approximately between 10 and 20 in simulations using the modified model (Section~\ref{subsec:modified_model}).
Therefore, in both the models for dendritic keratitis (Section~\ref{subsec:Model_dendritic_keratitis}) and the pseudodendritic keratitis model (Section~\ref{subsec:modified_model}), only patterns with central node distance map values ranging from 10 to 20 were included in the analysis.

For each branch, a pair of a Root value corresponding to the root of the branch and a Tip value corresponding to the terminal end was extracted from the sequence of distance map values and used for analysis. 

\begin{figure}
    \centering
    \includegraphics[width=1\linewidth]{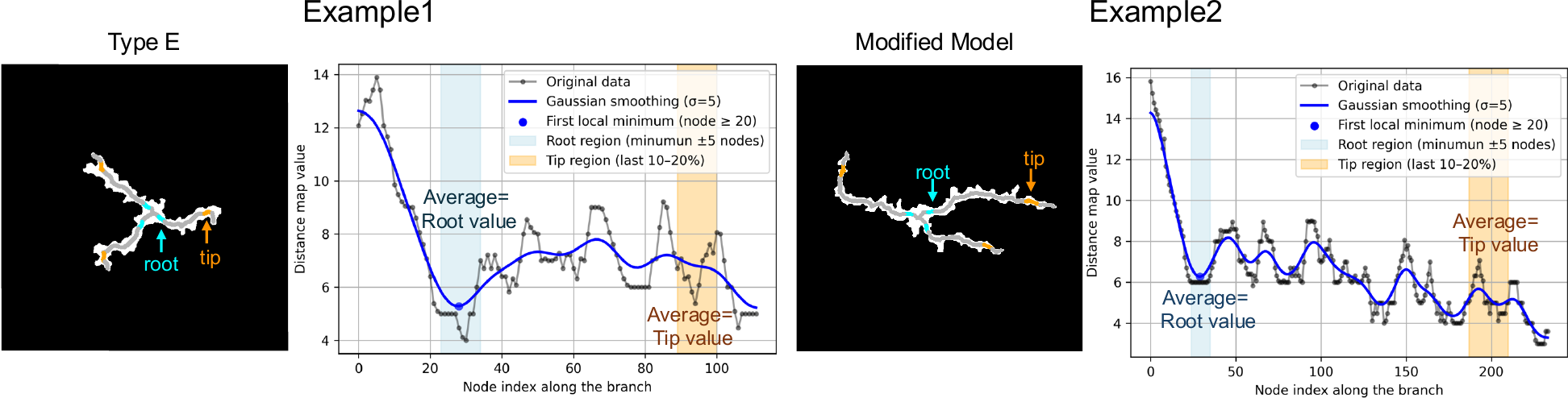}
    \caption{
    Definition of Root and Tip values for branch distance mapping. \textbf{For each branch, the distance map profile along the skeleton was first smoothed using a
one-dimensional Gaussian filter ($\sigma = 5$), and the Root location was identified as
the first local minimum appearing after the 20th skeleton node from the central node.
The Root value was defined as the average of the original distance map values within a
window of $\pm 5$ skeleton nodes around this location.
The Tip value was defined as the average of distance map values within a region located
between 10\% and 20\% from the terminal end of the branch, relative to the total branch
length.}
    }
    \label{fig:Analysis_method}
\end{figure}

\textbf{
To robustly identify the Root location, the distance map profile along each branch was
first smoothed using a one-dimensional Gaussian filter ($\sigma = 5$), and the first
local minimum appearing after the 20th skeleton node from the central node was detected.
Restricting the search to nodes beyond the 20th position was intended to avoid selecting
spurious minima arising from excessively thick proximal regions or from regions before
the branch was fully established.
The Root value was then defined as the average of the original (unsmoothed) distance map
values within a window of $\pm 5$ skeleton nodes around this local minimum
(Fig.~\ref{fig:Analysis_method}).
The Tip value was defined as the average of distance map values within a region located
between 10\% and 20\% from the terminal end of the branch, measured relative to the total
branch length.
}

The immediate terminal region was excluded from the Tip value calculation to avoid the influence of artificially low distance map values caused by proximity to the image boundary.

The normality of the difference between Root and Tip values was assessed using the Shapiro–Wilk test.  
If normality was confirmed, a \textbf{one-sided} paired t-test was performed, otherwise, \textbf{a one-sided} Wilcoxon signed-rank test was used\textbf{, based on the expected direction of the
difference for each model}.  
Based on the sample size and effect size, the statistical power was calculated.  
All statistical analyses were conducted using the \texttt{scipy} library.

\section{Results}
\subsection{Numerical simulation results}
\subsubsection{Model for the Formation of Dendritic Keratitis}
\begin{figure}
    \centering    \includegraphics[width=1.0\linewidth]{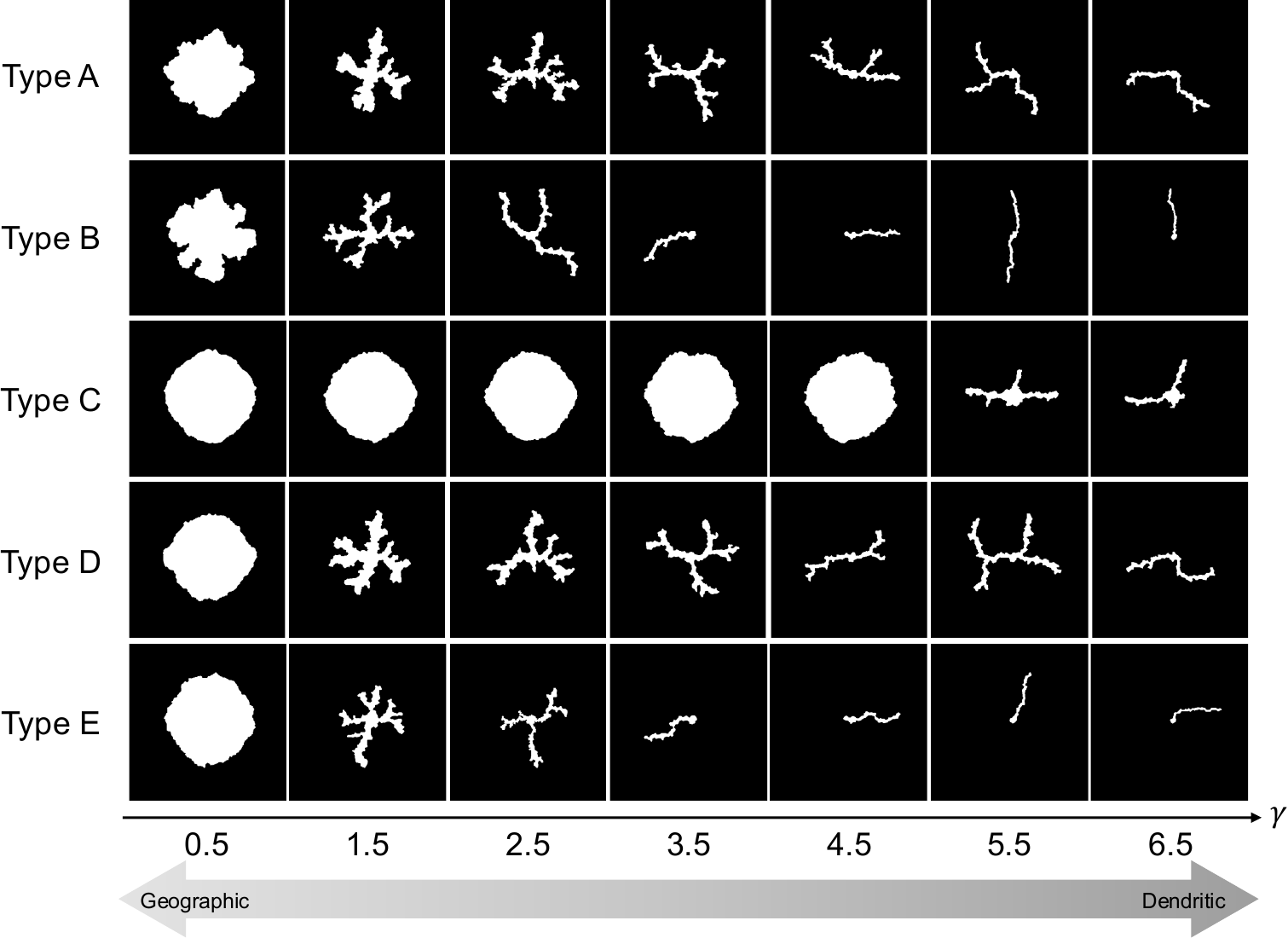}
    \caption{Numerical simulation results obtained by varying the cytokine secretion rate $\gamma$. The horizontal axis represents different values of $\gamma$ (ranging from 0.5 to 6.5), while the vertical axis corresponds to different cytokine-producing regions (Type A to E).  
    Lower $\gamma$ values resulted in geographic patterns, whereas higher values led to dendritic patterns.  
    Among dendritic patterns, lower $\gamma$ values produced more branched and thicker structures, while higher values yielded fewer branches and thinner extensions.
     All images are centered and cropped to highlight terminal morphology. 
}
    \label{fig:simulation_results}
\end{figure}

Numerical simulations were performed based on the aforementioned model (Section~\ref{subsec:Model_dendritic_keratitis}).  
Both geographic and dendritic patterns were generated regardless of Type A to E (corresponding to the cytokine-producing region $S$, Equation~\ref{eq:S_A}-\ref{eq:S_E}, Fig.~\ref{fig:fig_S}), depending on the parameter settings (Fig. \ref{fig:simulation_results}).  
Figure~\ref{fig:simulation_results} shows simulation results obtained by varying the cytokine secretion rate $\gamma$.  
Lower secretion rates resulted in geographic patterns, whereas higher rates led to dendritic patterns.  
Among the dendritic patterns, lower cytokine secretion induced more branched and thicker structures, while higher secretion produced fewer branches and thinner extensions. \textbf{Supplementary Figure~S2 provides the corresponding raw grid images, and 
Supplementary Movies~S1–S5 show the temporal evolution of these patterns.}

\begin{figure}
    \centering
    \includegraphics[width=1\linewidth]{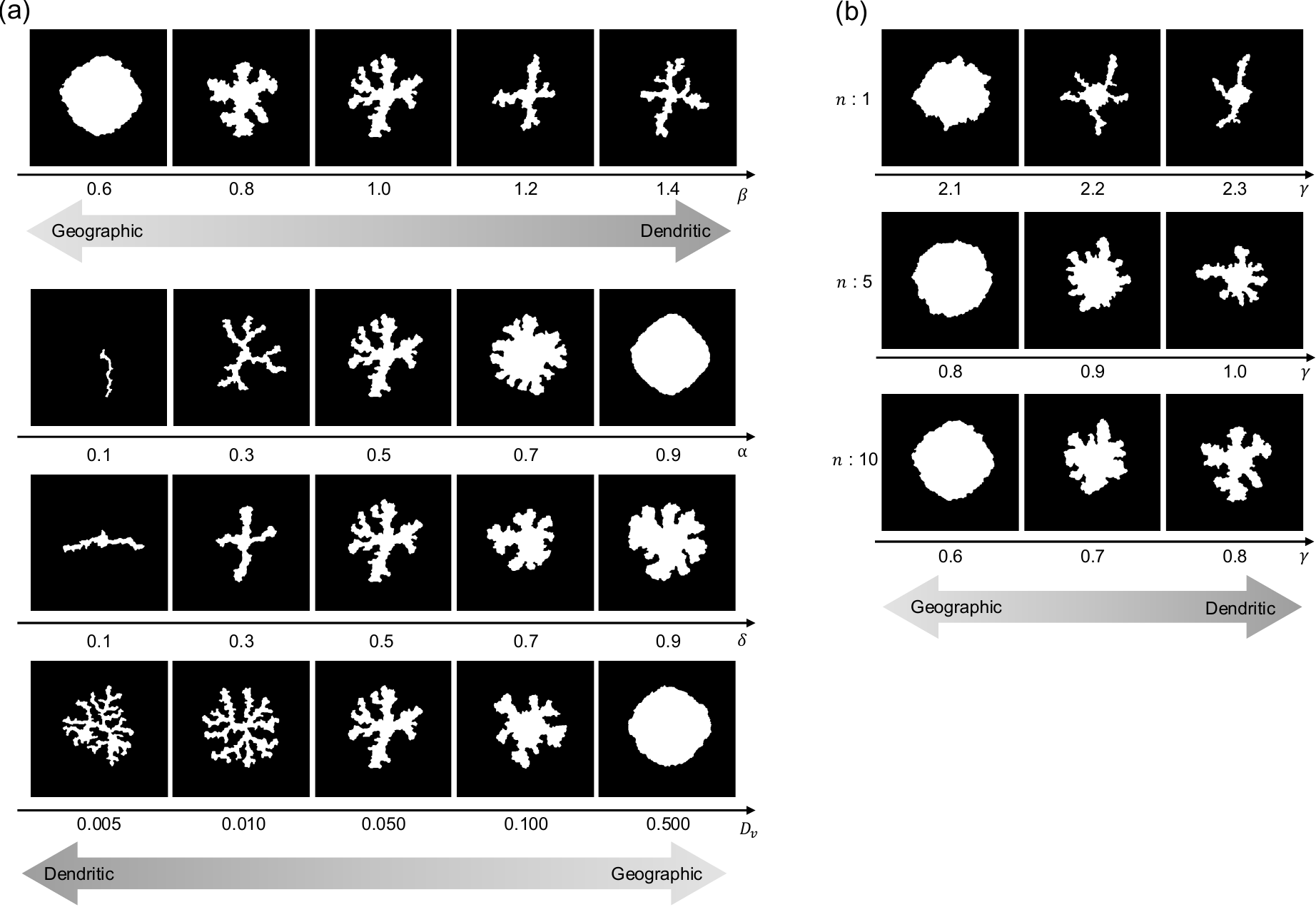}
    \caption{
    Simulation results of Type E under variations of parameters.  All images are centered and cropped to highlight terminal morphology. 
     (a) Simulation results showing the effect of varying $\alpha$ (\textbf{cytokine-independent net infection drive}), $\beta$ (cytokine-dependent infection inhibition), $\delta$ (cytokine degradation rate), and $ D_v $ (difference in diffusion coefficients), compared to the conditions used in Fig.~\ref{fig:simulation_results}.  
    \textbf{Increases in $\beta$ produced trends similar to increased $\gamma$, resulting in a transition from geographic to dendritic patterns, while increased $\alpha$ and $\delta$ showed the opposite effect}.  
    Larger values of $ D_v $ led to larger pattern size.  
    (b) Simulation results with varying the width $n$ of the cytokine-producing region and the parameter $\gamma$, which transitions from geographic to dendritic morphology.
    When $n = 1$, amoeboid and thin dendritic structures coexisted within a single simulation, resembling dendritic tails in geographic keratitis.  
    Such coexistence was not observed at higher $ n $ values.
    }
    \label{fig:other_parameters}
\end{figure}

In addition to variations in $\gamma$ (cytokine secretion rate), we examined the effects of other model parameters.  
$\beta$ (infection inhibition coefficient dependent on cytokines, e.g., cellular sensitivity to cytokines) exhibited trends similar to that of $\gamma$.  
In contrast, \textbf{both $\alpha$ (cytokine-independent net infection drive)} and $\delta$ (cytokine degradation rate) showed the opposite trend.  
Regarding $D_v$ (the diffusion coefficients), smaller values resulted in smaller pattern sizes, whereas larger values led to larger pattern sizes (Fig. \ref{fig:other_parameters}a).

In Type D and E models, the cytokine-producing region $S$ was defined by a width parameter $n$.  
When $n$ was small, changes in parameters $\alpha$ through $\delta$ led to abrupt variations in pattern size within a single simulation result.  
For $n=1$, amoeboid structures coexisted with thin structures, resembling dendritic tails observed in clinical cases of geographic keratitis.  
In contrast, when $n$ was large, such abrupt changes in pattern size within a single simulation were not observed (Fig.~\ref{fig:other_parameters}b).

\begin{figure}
    \centering
    \includegraphics[width=1\linewidth]{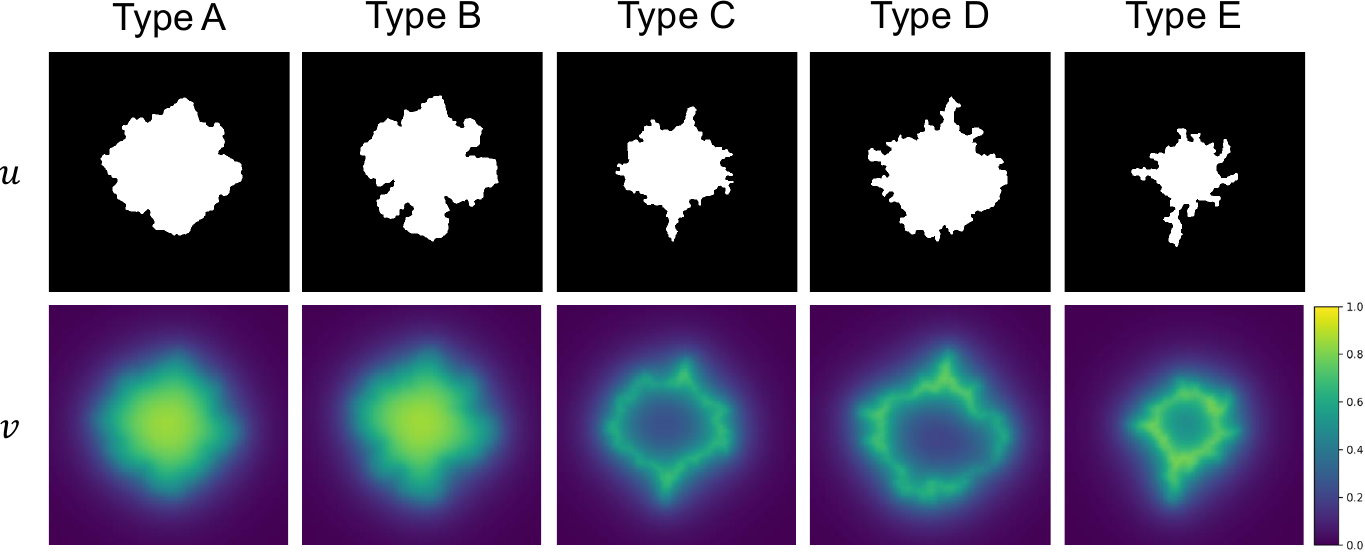}
    \caption{
    Simulation results in the parameter region where the lesion pattern transitions from geographic to dendritic.  
    The top row shows $u$; the lesion pattern, and the bottom row shows $v$; the corresponding cytokine concentration distribution.  
    All simulations were performed with parameters set to $\alpha = 0.5$, $\beta = 1.0$, $\delta = 0.5$, and $D_v = 0.05$, as listed in Table~\ref{tab:simulation_parameters}.  
    For Type D and Type E, the width of the cytokine-producing region was set to $n = 1$.  
    In Types A and B, cytokine concentrations were highest at the lesion center, whereas in Types C to E, concentrations were low within the lesion and elevated only at the periphery. All images are
centered and cropped to highlight terminal morphology. }
    \label{fig:cytokine_Dendritic_tail}
\end{figure}

In addition to Type E, we investigated whether dendritic tails form across all pattern types (Type A to E) in the parameter regions where patterns transition from geographic to dendritic.  
Dendritic tails were not observed in Type A or B, whereas they were formed in Types C to E; for Type D and E, this occurred only when the width parameter $n$ was small.  
Visualization of cytokine concentrations revealed that in geographic lesions formed under Type A and B conditions, cytokine levels were elevated at the lesion center.  
In contrast, for Types C to E, cytokine concentrations were low within the lesion and high only at the margins (Fig. \ref{fig:cytokine_Dendritic_tail}).

\subsubsection{Modified Model for Formation of Pseudodendritic Keratitis}
\begin{figure}
    \centering
    \includegraphics[width=1\linewidth]{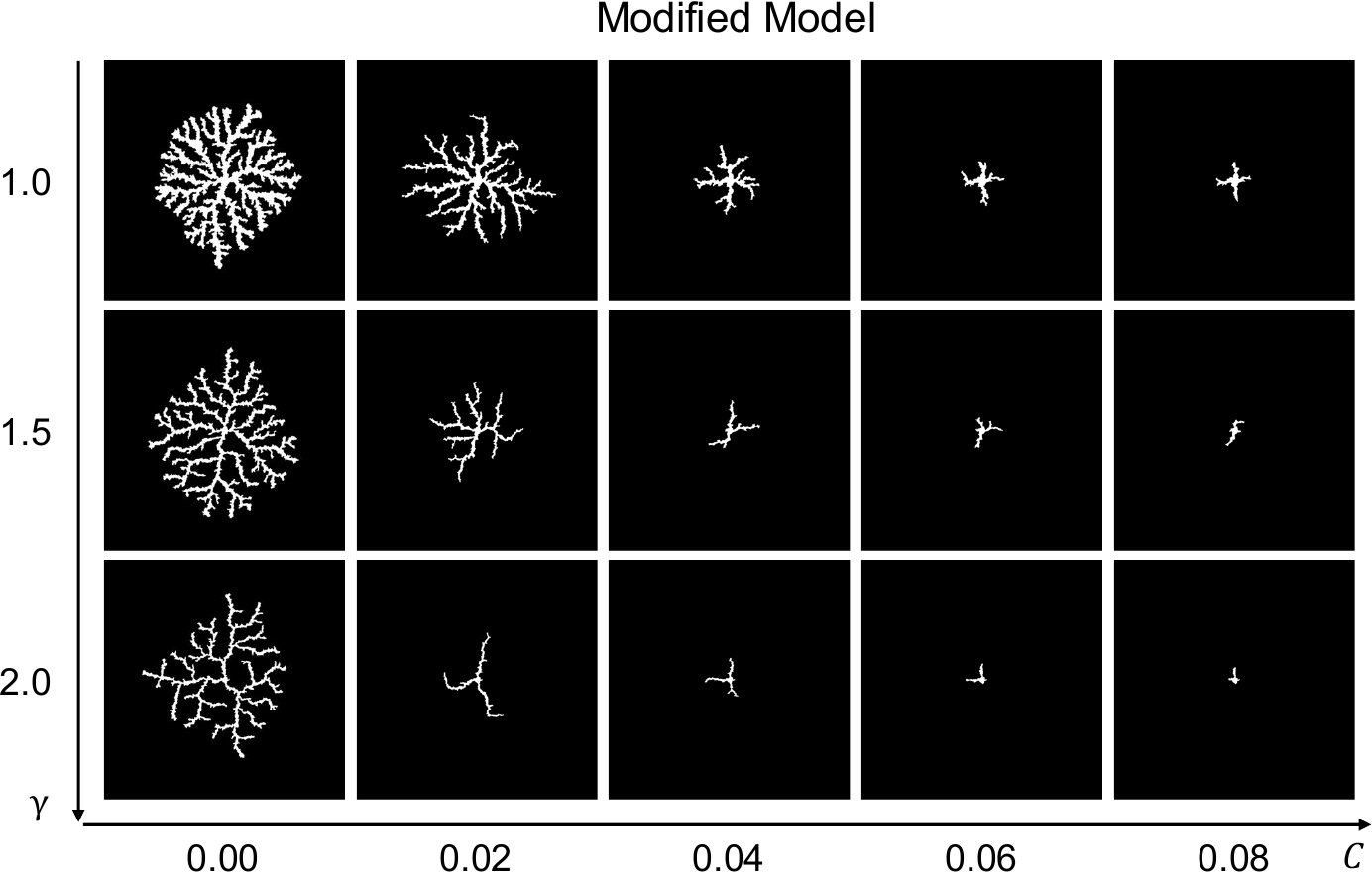}
    \caption{
    Simulation results from the modified model showing the effect of $C$; varying cytokine secretion rate at the limbus.  
    The horizontal axis represents $C$ (from \textbf{0.00 to 0.08}), and the vertical axis corresponds to $\gamma$ (\textbf{1.0, 1.5, 2.0}).  
    When $C$ was small, dendritic lesions with terminal bulbs were observed.  
    As $C$ increased, terminal bulbs were no longer formed and the lesion tips became tapered.  
    The simulations were configured to run for up to 10,000 iterations, or to terminate earlier if the lesion reached 90\% of the image width. 
    \textbf{For $C = 0.00$, the simulation terminated before reaching 10,000 iterations when the lesion reached this threshold: at 1,318 iterations for $\gamma = 1.0$, 1,693 iterations for $\gamma = 1.5$, and 1,851 iterations for $\gamma = 2.0$. }
    In contrast, for all cases with $C > 0.00$, the lesion did not reach 90\% of the image width, and simulations continued until the maximum 10,000 iterations were completed.
    }
    \label{fig:VZV_simulation_results}
\end{figure}

In the modified model for pseudodendritic keratitis (Section~\ref{subsec:modified_model}), when $C$ (the cytokine secretion rate at the limbus) was small, it had minimal effect on the lesion morphology.  
However, when $C$ was moderately increased, no terminal bulbs formed, and the terminal ends of the lesion instead became tapered (Fig. \ref{fig:VZV_simulation_results}).  
With further increases in $C$, lesion expansion was inhibited entirely, and the lesion remained in a small, undeveloped state. \textbf{The temporal evolution and cytokine concentration dynamics of the modified model are provided in Supplementary Movie~S6.}

\subsection{Image analysis}
\begin{figure}
    \centering
    \includegraphics[width=1\linewidth]{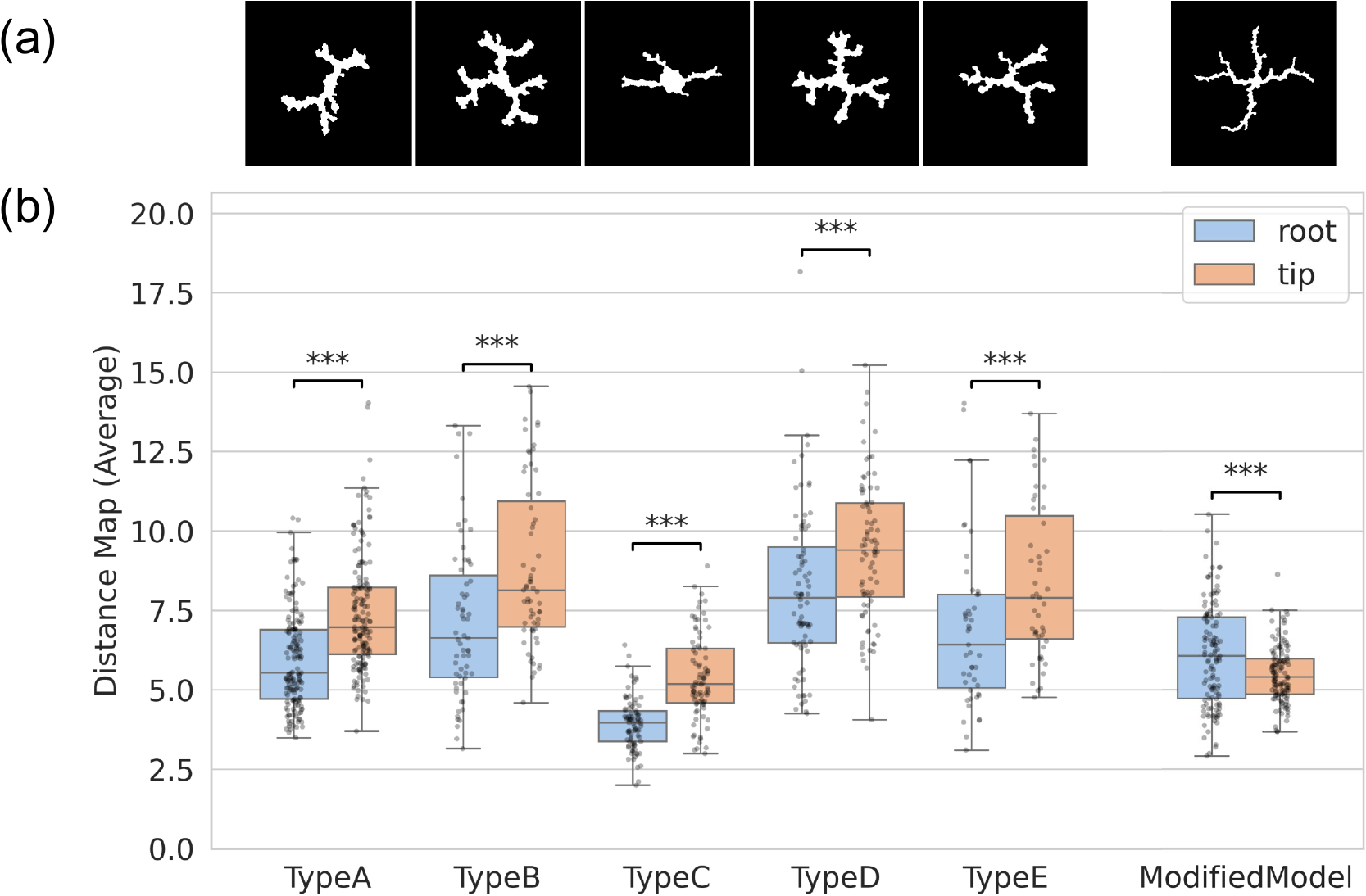}
    \caption{Terminal bulb analysis. 
    Qualitative and quantitative analysis of terminal bulb formation. 
    (a) Representative branch images from each model. All images are centered and cropped to highlight terminal morphology. Types A to E show bulbous, rounded terminal ends, whereas the modified model exhibits tapered, pointed terminal ends.  
    For Types D and E, the width parameter was set to $ n = 10 $. $\gamma$, the cytokine secretion rates used were A: 3.0, B: 2.5, C: 6.5, D: 3.0, E: 2.0, and for the modified model (corresponding to VZV): 4.5 with $ C = 0.003 $.  
    (b) Box plots comparing Root and Tip values of the distance map for each model. In Types A to E, Tip values were significantly greater than Root values, quantitatively confirming the presence of terminal bulbs.  
    In contrast, the modified model showed significantly lower Tip values than Root values, indicating the absence of bulb formation. *** denotes statistical significance with $ p < 0.001 $.
    }
    \label{fig:box_plots}
\end{figure}

Statistical analysis was performed to evaluate the formation of terminal bulbs.  
Box plots comparing Root and Tip values for each branch were generated for Types A to E and the modified model.  
In all of Types A to E, the Tip values were significantly higher than the Root values, quantitatively confirming the presence of terminal bulbs \textbf{(Fig.~\ref{fig:box_plots} and Fig. S3)}.  
In contrast, in the modified model, the Tip values were significantly lower than the Root values, indicating the absence of terminal bulbs.  
These quantitative results were consistent with visual inspection.

The sample size, effect size, statistical power, $p$-value and statistical method used for each analysis are summarized in \textbf {Table.S1}. \textbf{We additionally applied the same root–tip measurement procedure to clinical images of dendritic and pseudodendritic keratitis. The clinical data showed the same qualitative relationship as in the simulations, with HSV cases exhibiting root < tip and the VZV case showing root > tip (Supplementary Figure S5 and S6).}

\begin{figure}
    \centering
    \includegraphics[width=1.0\linewidth]{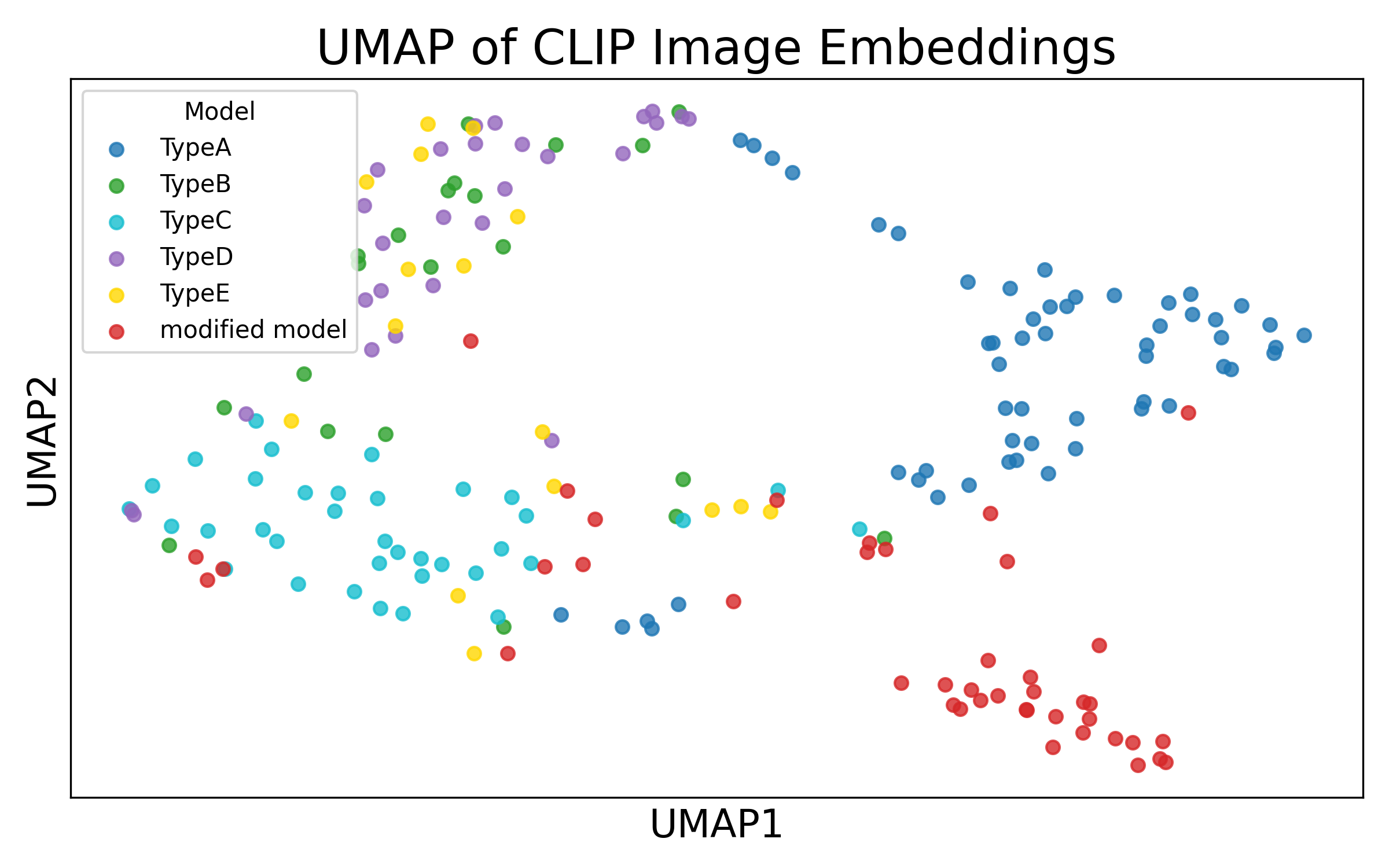}
    \caption{Two-dimensional UMAP projection of image embeddings extracted by CLIP (RN50).  
    \textbf{The modified model (red) occupies a partially separated region of the embedding space,
    indicating that its lesion patterns differ systematically from those generated by the cytokine-dependent models (Types A–E).
    The other models form an overall contiguous cluster, although substructure is present, with each model showing local separation within the shared manifold.
    These results are consistent with the qualitative and quantitative differences in terminal-end morphology observed between the modified model and the cytokine-dependent models.}
    }
    \label{fig:CLIP_UMAP}
\end{figure}

Machine learning-based analysis was also conducted to compare the simulation results from the two models\cite{Hishinuma2025}.  
The latent space extracted by Contrastive Language–Image Pre-Training (CLIP) with a ResNet50 (RN50) image encoder was visualized using UMAP (Fig.~\ref{fig:CLIP_UMAP}).  
In the previous subsection, Types A to E exhibited visually bulbous terminal ends and statistically higher Root values than Tip values, whereas the modified model showed tapered tips and Tip values significantly greater than Root values. 
\textbf{To ensure consistency with the statistical analysis, only simulation images whose central distance map values fell within the 10–20 range were included in the UMAP embedding.}
This distinction was also supported by the machine learning analysis: \textbf{the modified model (red) occupied a partially separated region in the UMAP embedding space, indicating that its lesion patterns differ systematically from those generated by the cytokine-dependent models.
The other models (Types A–E) formed an overall contiguous cluster, although substructure was present, with each model showing local separation within the shared manifold.
These observations support the conclusion that the modified model generates a distinct class of lesion shapes, consistent with its cytokine-independent mechanism of tip narrowing.}

\section{Discussion}
\subsection{Research summary}
In this study, we developed a mathematical model to reproduce the pattern formation observed in dendritic keratitis. \textbf{Our model incorporated a cell-to-cell infection mechanism, which is consistent with experimental findings that dendritic spread occurs in organotypically cultured corneas but not in isolated corneal cell cultures lacking tight intercellular junctions \cite{Thakkar2017}. 
This supports the assumption that direct viral transmission between adjacent epithelial cells is essential for reproducing the dendritic morphology observed in vivo. In addition,}consistent with the previously proposed hypothesis that immune responses play a significant role in shaping the morphology of dendritic keratitis \cite{Thakkar2017}, our numerical simulations demonstrated that when the effects of infection-suppressive cytokines were strong, dendritic patterns were generated.

\textbf{This finding supports the notion that the secretion and spatial distribution of cytokines from infected epithelial cells are likely to regulate the direction and rate of infection, thereby contributing to the morphology of dendritic keratitis.}

Conversely, when the impact of cytokines was weak, the lesions expanded and adopted a more geographic morphology. Intuitively, when the effect of cytokine is significant, protruded lesions can grow faster than other regions since they are away from other cytokine-producing regions, resulting in the formation of branches. On the other hand, when the effect of cytokine is weak, the growth speeds of protruded and concaved regions are similar, resulting in a geographic pattern. 

\subsection{Mechanism of dendritic tail formation}
The dendritic tail, a characteristic feature observed in geographic keratitis, was also reproduced in our numerical simulations by narrowing the width of the cytokine-producing region. Biologically, it is known that infected cells can undergo necrosis or desquamation, resulting in a distribution where infected cells remain predominantly at the lesion margin. Our results suggest that the dendritic tail may form under such conditions. Currently, we do not understand the exact mechanism of this pattern formation since analytical methods are limited to this discrete and stochastic system. 

\subsection{Terminal bulb formation}
In the model replicating dendritic keratitis, both visual inspection and statistical analysis demonstrated the formation of terminal bulb-like structures. In contrast, the modified model did not produce terminal bulbs. Examining the differences between these models revealed that, in the dendritic keratitis model, cytokine production was limited to the periphery of the already established infection lesions. As a result, it took time for the cytokine concentration to rise at the lesion periphery, and diffusion led to a further decrease in cytokine levels near the terminal ends of the lesion. This maintained lower cytokine concentrations at the terminal ends of the lesion, which helped preserve the pattern size and allowed the formation of terminal bulb-like structures.

In the pseudodendritic keratitis model, however, cytokines were produced at the boundary of the computational domain and diffused rapidly, \textbf{causing the cytokine concentration near the terminal ends of the lesion to increase over time}. Consequently, the pattern size of lesions formed later during infection became smaller than that of lesions formed earlier, leading to a tapered lesion morphology. This may correspond to the more severe inflammation observed in pseudodendritic keratitis caused by VZV than in dendritic keratitis. \textbf{As illustrated in Supplementary Movie~S7, tapered morphologies can also emerge in a multi-seed setting where neighboring lesions provide inhibitory cytokines. Although this condition does not directly correspond to clinical observations, it shows that an external cytokine gradient is sufficient to generate a tapered advancing tip. This supports the conclusion that tapered lesion shapes require cytokine input originating outside the lesion itself.}

\textbf{\subsection{Limitation of study}}\label{limitation}

\textbf{The present model focuses on the epithelial layer and assumes that dendritic lesion morphogenesis arises primarily from cell-to-cell viral transmission modulated by cytokine-mediated immune responses.
This framework does not incorporate stromal cells, the basement membrane, or the multilayered structure of the corneal epithelium, all of which are present in ex vivo or in vivo corneal tissue and may influence epithelial infection dynamics.
However, since the corneal epithelium is less than a hundred micrometers thick \cite{Wu2014} compared with its ~12 mm diameter, the effects of vertical (three-dimensional) structure are likely to be minor relative to lateral spreading processes in the epithelial plane.}

\textbf{\subsection{Future directions}}
\textbf{In this study, we investigated the mechanisms of dendritic keratitis pattern formation from a mathematical perspective. Based on these findings, it is expected that the addition of cytokines in ex vivo experiments may also reproduce these patterns biologically, providing further validation for the model.}

\section{Code availability}
All analysis and simulation code has been made publicly available at:\\
\url{https://github.com/marimasunaga/herpes_keratitis_model}

\section{Acknowledgment}
Computations were partially performed on the NIG supercomputer
at ROIS National Institute of Genetics.

\section{Statements and Declarations}
The authors have no relevant financial or non-financial interests to disclose.

\bibliographystyle{unsrt} 
\bibliography{dendritic_keratitis} 

\newpage

\let\oldsubsection\subsection
\let\oldthesubsection\thesubsection
\setcounter{subsection}{0}
\renewcommand{\thesubsection}{\arabic{subsection}}


\newcounter{appsubsection}[section]
\renewcommand\theappsubsection{S\arabic{appsubsection}}

\makeatletter
\renewcommand\subsection[1]{
  \stepcounter{appsubsection}
  \par\vspace{3ex \@plus .2ex}
  \noindent\textbf{S\theappsubsection\ #1}
  \par\vspace{1ex \@plus .2ex}
}
\makeatother

\newcounter{appsubsubsection}[subsection]
\renewcommand\theappsubsubsection{\theappsubsection.\arabic{appsubsubsection}}

\makeatletter
\renewcommand\subsection[1]{
  \refstepcounter{appsubsection}
  \par\vspace{1ex \@plus .2ex}
  \noindent\textbf{\theappsubsection\ #1}
  \par\vspace{1ex \@plus .2ex}
}
\makeatother

\setcounter{appsubsection}{0}
\renewcommand\theappsubsection{}

\end{document}